\documentclass[doublecol]{epl2} 

\newcommand{\rb}[1]{\raisebox{1.5ex}[0mm]{#1}}
\title{New fitting scheme to obtain effective potential from Car-Parrinello molecular 
dynamics simulations~: Application to silica}
\shorttitle{New effective potential for silica}

\author{A. Carr\'e\inst{1,2} \and J. Horbach\inst{1,3} \and S. Ispas\inst{2} 
\and W. Kob\inst{2}}
\shortauthor{A. Carr\'e \etal}

\institute{
\inst{1} Institut f\"ur Physik, Johannes Gutenberg--Universit\"at Mainz, 
Staudinger Weg 7, 55099 Mainz, Germany\\
\inst{2} Laboratoire des Collo\"\i des, Verres et Nanomat\'eriaux, 
Universit\'e Montpellier II and CNRS UMR 5587, 34095 Montpellier, France\\
\inst{3} Institut f\"ur Materialphysik im Weltraum, Deutsches Zentrum f\"ur 
Luft-- und Raumfahrt (DLR), 51170 K\"oln, Germany
}


\abstract{
A fitting scheme is proposed to obtain effective potentials from
Car-Parrinello molecular dynamics (CPMD) simulations.  It is used
to parameterize a new pair potential for silica.
MD simulations with this new  potential are done to determine structural
and dynamic properties and to compare these properties to those obtained 
from CPMD and a MD simulation using the so-called BKS potential.
The new potential reproduces accurately the liquid structure generated by
the CPMD trajectories, the experimental activation energies
for the self-diffusion constants and the experimental density of
amorphous silica.  Also lattice parameters and elastic constants of
$\alpha-$quartz are well-reproduced, showing the transferability of
the new potential.}

\begin{document}

\maketitle

\section{Introduction}

One of the central issues of atomistic computer simulations is to develop
effective potentials \cite{finnis03} that model the interactions between
ions by some functional form without taking into account explicitly
the electronic degrees of freedom.  The first step to obtain such a
potential for a given system is to decide its functional form. In
terms of computational efficiency, an optimal ansatz is provided by
a pair potential model.  However, for a realistic modelling it might
be often necessary to include many-body terms to describe, e.g.,
three-body interactions or polarizability effects.  In a second step,
the free parameters of the potential function have to be fitted to
experimental data and/or to {\it ab initio} calculations \cite{BKS,tangney02}. A relatively
new approach is to use {\it ab initio} molecular dynamics simulations
such as Car-Parrinello molecular dynamics (CPMD) \cite{car85} for the
parametrization of effective potentials. The latter simulation technique
has been proven very successful to realistically describe a large number
of atomistic systems, among them network forming systems such as silica
\cite{pasquarello98,sarnthein95,umari07,Benoit_Ispas00,Benoit_Ispas01,Benoit_Kob02}.
Nowadays, systems of 100 to 200 particles can be simulated on a time
scale of several tens of picoseconds by CPMD. 

In this Letter, we aim at developing a fitting scheme for 
the parametrization of effective potentials from CPMD trajectories
and we address this issue to the case of liquid silica.
A widely-used method for the parametrization of potentials is the
so-called force matching procedure \cite{ercolessi94} where one tries to match the CPMD
forces on the atoms with those resulting from the classical potential.
However, we have found that, in the case of silica, the force matching
method fails (at least if a pair potential model is considered) \cite{Carre_PhD07}:
Although the CPMD forces are well reproduced by the resulting effective
(pair) potential, the structure is very different from that obtained by
CPMD. The main aim of this work is to develop a different fitting scheme
where one tries to match as closely as possible the atomistic structure,
as obtained from the classical MD using the effective potential, with
that of the CPMD.

Since silica is of great importance for technological
applications, geosciences and the theory of the glass transition,
the development of effective potentials for silica has a long history
\cite{woodcock76,tsuneyuki88,jackson88,vashishta90,BKS,wilson93,wilson96,tangney02}.
Surprisingly, various studies have shown that a simple
pair potential, the so-called BKS potential \cite{BKS}, is
able to give a quite accurate description of amorphous silica
\cite{Benoit_Ispas00,vollmayr96,horbach99,horbach99b,herzbach05}.  The
parametrization of this potential was based on Hartree-Fock calculations
of a single SiO$_4$ tetrahedron (saturated by four hydrogen atoms),
considering also experimental elastic constants of $\alpha-$quartz in
the fitting procedure.

Unsurprisingly, the BKS model has also several deficiencies. For
instance, the equation of state, as far as it is known experimentally,
is not reproduced well \cite{tangney02}. One could argue that one has
to include many body effects to overcome the deficiencies of the BKS
potential. However, as shown in Ref.~\cite{herzbach05}, more complicated
potentials that include polarizability effects or fluctuating charges
do not provide necessarily a more realistic description of silica.

Here, we demonstrate that one can improve the accuracy of a pair
potential for amorphous silica.
The idea
of our approach is to match the partial pair correlation functions,
as obtained from the effective potential, with those obtained from
a CPMD simulation. As demonstrated below, the new potential yields
an accurate description of amorphous silica with respect to density,
structure and diffusion dynamics. Also properties of $\alpha-$quartz
are well reproduced by the new potential.  The fitting scheme proposed
in this work can be also used for other systems and it is not
restricted to the parametrization of pair potentials.

\section{Simulation details and fitting methodology}
\subsection{Ab initio simulations\label{AIS}} 
CPMD simulations \cite{CPMD} were done for a system of 38 SiO$_2$ units
(114 atoms) in a cubic box of size $L=11.982$\,{\AA}, corresponding to
the experimental density of $2.2$\,g/cm$^3$ \cite{mazurin}. Periodic
boundary conditions are used in all three spatial directions. The
electronic structure was described by the local density approximation
(LDA) in the framework of density functional theory \cite{KS}. Norm-conserving 
pseudopotentials were used for silicon \cite{BHS} and oxygen
\cite{MT}, and the plane-wave basis sets were expanded up to
an energy cut-off of 70\,Ry.  The choice of pseudopotentials,
exchange and correlation functionals and the plane-wave cutoff are
justified by previous CPMD studies carried out on amorphous SiO$_2$
\cite{Benoit_Ispas00,Benoit_Ispas01,Benoit_Kob02}.

A time step of 3\,a.u.~(0.0725\,fs) and a fictitious electronic mass of
600\,a.u. were used to integrate the equations of motion. Temperature is
kept constant by applying Nos\'e-Hoover chains for each ionic degree of
freedom \cite{martyna92,martyna96} as well as for the electronic degrees
of freedom to counterbalance the energy flow from ions to electrons
\cite{tuckerman94}. The parameters used for the Nos\'e-Hoover chains can
be found in previous publications \cite{Benoit_Ispas00,Benoit_Ispas01}.

The starting configuration for the CPMD was generated by
MD simulations with the BKS potential \cite{BKS}, following the
same methodology as in Refs.~\cite{Benoit_Ispas00,Benoit_Ispas01}.
Then, a CPMD run over about 20\,ps at $T=3600$\,K was 
performed from which we used the last 16.5\,ps for the analysis.

\subsection{Fitting procedure}
The next step was to obtain an effective potential from the CPMD data.
To this end, we aimed at parameterizing a given potential model
such that the partial pair correlation functions $g_{\alpha \beta }(r)$
(${\alpha \beta \in \{\rm{Si,O}\}}$) \cite{glassbook}, as calculated from
the CPMD trajectories, were reproduced by the MD simulations using the new
potential.

As an ansatz for the effective potential we used the same functional
form as the one used for the BKS potential,
\begin{equation} 
  u_{\alpha \beta}(r) =
     \frac{q_{\alpha} q_{\beta} e^2}{r} +
      A_{\alpha \beta} \exp\left( -B_{\alpha \beta} r\right) -
      \frac{C_{\alpha\beta}}{r^6} ,
   \label{BPOTENTIAL}
\end{equation} 
where $r$ is the distance between an ion of type $\alpha$ and an ion
of type $\beta$ ($\alpha, \beta = {\rm Si, O}$) and $e$ is the elementary
charge. We consider the parameters appearing in Eq.~(\ref{BPOTENTIAL}) as elements
of the vector ${\bf \xi}=(\xi_k,\, k=1,\, 2,\ldots 10) = (q_{\rm{Si}},
A_{\alpha\beta}, B_{\alpha\beta}, C_{\alpha\beta}, \alpha, \beta \in
\{\rm{Si,O}\})$. The effective charge for the oxygen atoms
does not appear in the $\xi$ vector since, due to the charge neutrality
requirement, it is directly related to the silicon charge via $q_{\rm
O}=-q_{\rm Si}/2$.  Forces and energies that correspond to the long-ranged
Coulomb terms in (\ref{BPOTENTIAL}) have been computed by Ewald sums
\cite{allen}, those for the short-range part in (\ref{BPOTENTIAL}) have
been truncated and shifted at $6.5$\,{\AA}. In addition a smoothening
function was used for the short-range part to avoid a drift of the total
energy in microcanonical MD runs. The details of this function can be
found in a recent MD study with the BKS model \cite{carre07}.

The cost function $\chi^2$ that measures the difference between the
CPMD data and the fitting model is based on the pair correlation functions
weighted by the distance $r$, $rg_{\alpha \beta}(r)$,
\begin{equation}
\chi^2  =  \int_{0}^{L/2} \lambda^2(r)dr ,  
\label{partial_chi2}
\end{equation}
where $\lambda^2 (r)$ is defined as follows:
\begin{equation}
\lambda^2 (r)   =  \sum_{\alpha, \, \beta={\rm Si, O}} \left[
r g_{\alpha  \beta}^{\rm{CPMD}}(r)-
       r g_{\alpha \beta}^{\rm{CHIK}}(r;\xi)
\right]^2 .
\label{rgr_criterion}
\end{equation}
Here, the superscripts CPMD and CHIK indicate respectively the use
of CPMD and MD with the new potential, that we call CHIK potential
in the following. The integral (\ref{partial_chi2}) was evaluated by
numerical integration. Note that a cost function $\chi^2$ in which the function
$r g_{\alpha\beta}(r)$ in Eq. (\ref{rgr_criterion}) is replaced by
$g_{\alpha\beta}(r)$ or $r^2 g_{\alpha\beta}(r)$ resulted in potentials
that were not accurate \cite{Carre_PhD07}. The reason for this is 
that $rg_{\alpha\beta}(r)$ makes a good compromise between weighting the
structural information at short distances, such as is done by the cost function
with $g_{\alpha\beta}(r)$, and long distances, as is done by the cost function
with $r^2 g_{\alpha\beta}(r)$.

\begin{table}
\begin{center}
\begin{tabular}{|cl|c|} \hline
\multicolumn{2}{|c|}{Parameter} & CHIK \\ \hline\hline
$q_{\rm{Si}}$   & (C)            & 1.910418     \\
$A_{\rm{OO}}$   & (eV)           & 659.595398   \\ 
$B_{\rm{OO}}$   & ({\AA}$^{-1}$) & 2.590066     \\
$C_{\rm{OO}}$   & (eV.{\AA}$^6$) & 26.836679    \\
$A_{\rm{SiO}}$  & (eV)           & 27029.419922 \\ 
$B_{\rm{SiO}}$  & ({\AA}$^{-1}$) & 5.158606    \\
$C_{\rm{SiO}}$  & (eV.{\AA}$^6$) & 148.099091   \\
$A_{\rm{SiSi}}$ & (eV)           & 3150.462646  \\
$B_{\rm{SiSi}}$ & ({\AA}$^{-1}$) & 2.851451     \\
$C_{\rm{SiSi}}$ & (eV.{\AA}$^6$) & 626.751953   \\ \hline
\end{tabular}
\caption{Parameters of the CHIK potential.
\label{RGR_PARAM}}
\end{center}
\end{table}
The optimal vector ${\mathbf \xi}$ of the potential parameters was determined by an
iterative Levenberg-Marquardt algorithm \cite{Numerical_Recipes}. After
each iteration step, ${\mathbf \xi}$ was updated leading to a
new guess for the potential. This potential was then used in MD
simulations yielding a new set of pair correlation functions $g_{\alpha
\beta}^{\rm{CHIK}}(r;\xi)$.  Recall that the Levenberg-Marquardt algorithm requires
the calculation of the derivatives of the $g_{\alpha\beta}$ with respect
to the potential parameters. These derivatives were
computed numerically by finite differences:
\begin{equation}
\frac{\partial g_{\alpha \beta}(r,\xi)}{\partial \xi_k}
=\lim_{\varepsilon_k \to 0}
\frac{g_{\alpha \beta}(r,\xi_k+\varepsilon_k)-g_{\alpha \beta}
(r,\xi_k-\varepsilon_k)}{2 \varepsilon_k}.
\label{eq_deriv}
\end{equation}
Here, small perturbations ($\pm \varepsilon_k$) of
a given parameter $\xi_k$ are considered, keeping all other parameters
constant.

MD simulations were performed to compute the
functions $g_{\alpha \beta}^{\rm{CHIK}}(r;\xi)$ and the derivatives
(\ref{eq_deriv}).  In these simulations, we considered a system of
1152 atoms in a cubic box of size $25.9$\,{\AA} (corresponding to
the density $\rho=2.2$\,g/cm$^3$). The simulations were
done in the $NVT$ ensemble at the temperature $T=3600$\,K. Temperature
was kept constant by coupling the system periodically to a stochastic
heat bath. The equations of motion were integrated by the velocity form
of the Verlet algorithm, using a time step of $1.0$\,fs. For an accurate
calculation of the functions $g_{\alpha \beta}^{\rm{CHIK}}(r;\xi)$,
runs over 10\,ps were required.

Following the methodology described above, convergence was obtained after
44 iterations yielding the potential parameters\footnote{As the Born-Mayer ansatz diverges
for very short distances, the following repulsive interaction has been
added to the potential ansatz (\ref{BPOTENTIAL}):
\begin{eqnarray*}
u^{\rm{cor}}_{\alpha \beta}(r) = \frac{D_{\alpha\beta}}{r^{24}} 
\nonumber
%
\end{eqnarray*}
with $D_{\rm{OO}}=113.0$\,eV.{\AA}$^{24}$, 
$D_{\rm{SiO}}=29.0$\,eV.{\AA}$^{24}$,
$D_{\rm{SiSi}}=3423200.0$\,eV.{\AA}$^{24}$.}
that are listed
in Tab.~\ref{RGR_PARAM}.  We note that these parameters are of the same
order of magnitude as the parameters of the BKS potential \cite{BKS}.
Thus, our model can be seen as a modification of the BKS potential
which yields already a good description of (amorphous) silica.

\subsection{MD simulations with the CHIK potential}
As a third step extensive MD simulations were performed with the
CHIK potential to investigate the temperature dependence of
structural and dynamic properties of the new model.  We used systems
of 1152 particles in a cubic box of size $L=25.9$\,\AA\, corresponding
to the density $\rho=2.2$\,g/cm$^3$.  Equilibration runs were done at
constant temperature (see above), using now a time step of 1.6\,fs. After
the equilibration, microcanonical production runs were done at the
temperatures 5200\,K, 4300\,K, 4000\,K, 3580\,K, 3250\,K, 3000\,K,
2900\,K, 2750\,K, 2580\,K, and 2440\,K.  At
$T=2440$\,K, each the equilibration and production runs lasted over 32
million time step corresponding to a real time of 52.3\,ns.  At each
temperature, results were averaged over 8 independent runs.  In order
to estimate the pressure, additional runs were done at $2230$\,K and
$2000$\,K. Here, the system was annealed for 15 million time steps.
Then the pressure was measured in subsequent runs over 5 million
time steps.

For comparison, also results from a previous MD simulation with the BKS
model are shown in the following. The details of this simulation can be
found in Ref.~\cite{horbach99}.

\section{Results}
\subsection{Static properties}
\begin{figure}
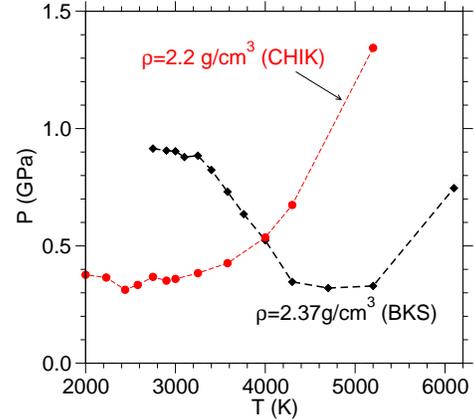

\onefigure[scale=0.31]{fig3.eps}
\caption{Temperature dependence of the pressure for the BKS model
(from Ref.~\cite{horbach99}) and the CHIK model, as indicated.}
\label{fig1}
\end{figure}
One of the peculiar properties of amorphous SiO$_2$ is the occurrence
of a density maximum at 1820\,K \cite{brueckner70}. Moreover, in a broad
temperature range from room temperature to temperatures above 2000\,K the
variations of the density around an average value of about 2.2\,g/cm$^3$
are smaller than 1\% \cite{mazurin}.  These features can also be
seen in the temperature dependence of the pressure since the pressure
shows a local  minimum  at the temperature where the density maximum
occurs. Both the BKS and the CHIK model show indeed a minimum pressure
at about 4800\,K and 2300\,K, respectively, see Fig.~\ref{fig1}. However,
the result for the CHIK model is in better agreement with experiment with
respect to the location of the minimum and the density. As can be inferred
from Fig.~\ref{fig1}, at $\rho=2.2$\,g/cm$^3$, 
a minimum value of about 0.35\,GPa
is found for the CHIK model. In contrast to this the BKS model has,
at a density of $\rho=2.2$\,g/cm$^3$, negative values for the pressure
if $T<7000$\,K \cite{vollmayr96}.

\begin{figure}
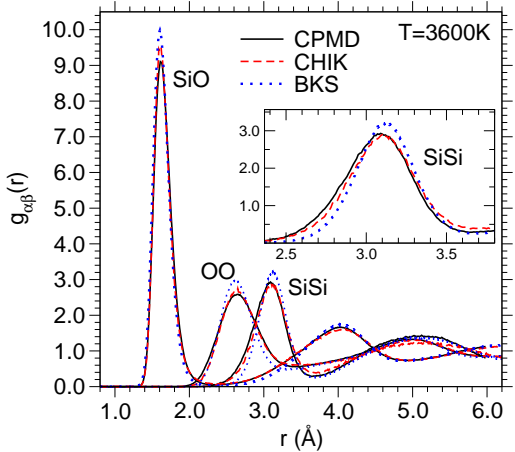

\onefigure[scale=0.33]{fig1.eps}
\caption{Partial pair correlation functions at $T=3600$\,K,
calculated from CPMD and classical MD using the
BKS and the CHIK potentials, as indicated. The inset shows an enlargement
of the first peak in $g_{\rm SiSi}(r)$.}
\label{fig2}
\end{figure}
Figure \ref{fig2} shows the
partial pair correlation functions at $T=3600$\,K, as calculated from
CPMD and classical MD using the BKS and the CHIK model. The
CHIK model yields good agreement with the CPMD results. The largest
differences are found in $g_{\rm SiSi}(r)$, but also in this case,
the CHIK potential leads to a better agreement with CPMD than the BKS
potential (see the inset in Fig.~\ref{fig2}).

\begin{figure}
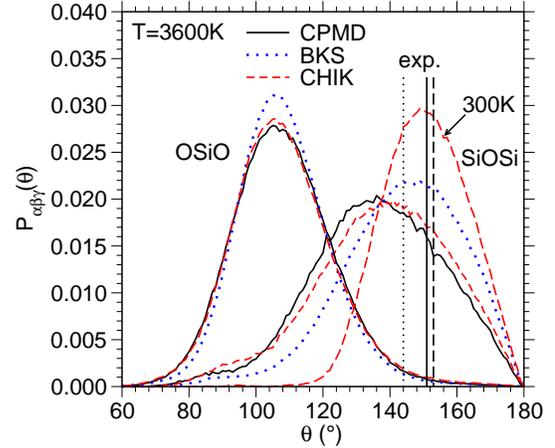

\onefigure[scale=0.33]{fig2.eps}
\caption{Angular distribution functions for the OSiO and SiOSi angle at 
$T=3600$\,K. For the CHIK model, the SiOSi distribution at $T=300$\,K is
also included. The vertical lines indicate experimental values from 
Refs.~\cite{mozzi69} (dotted line), \cite{mauri00} (solid line),
and \cite{pettifer88} (dashed line).}
\label{fig3}
\end{figure}
The partial pair correlation functions as obtained from the CPMD run
were used to parametrize the CHIK potential. That these functions are
reproduced well by the CHIK potential is thus not that surprising.
A less obvious test of the CHIK model is provided by considering
the distribution functions $P_{\alpha\beta\gamma}$ of the bond angles
($\alpha\beta\gamma={\rm OSiO, SiOSi}$) in comparison to CPMD.  As can be
inferred from Fig.~\ref{fig3}, the average intra-tetrahedral OSiO angle is
around $\theta=106^{\circ}$ at 3600\,K, both for CPMD and classical MD.
However, in the case of the BKS model the function $P_{\rm OSiO}$ is less
broad and exhibits a significantly higher amplitude than the CPMD and the
CHIK model.  Larger differences are seen in the distribution function
for the inter-tetrahedral SiOSi angle.  While at 3600\,K the BKS distribution
shows a maximum around $\theta=147^{\circ}$, the CHIK potential gives a
maximum around $\theta=141^{\circ}$, in better agreement with the CPMD
value around $\theta=136^{\circ}$.  

The distribution functions for the SiOSi angle exhibit also a
shoulder around $\theta=90^{\circ}$ revealing the emergence of
edge-sharing tetrahedra at the relatively high temperature $T=3600$\,K
\cite{vollmayr96}. In the BKS case, the latter shoulder has a lower
amplitude, indicating a smaller number of edge-sharing tetrahedra.
As for $P_{\rm OSiO}$, the BKS function for $P_{\rm SiOSi}$ is less broad
and its main peak has a higher amplitude. In this sense, the BKS model
leads to a less disordered structure than the CPMD and the CHIK model.
Also included in Fig.~\ref{fig3} is the SiOSi bond angle distribution at
300\,K (i.e.~in the glass), as obtained from the simulation with the CHIK
model (for the preparation of the glass samples, see below the discussion
on the calculation of the density of states).  In agreement
with experimental data \cite{mozzi69,mauri00,pettifer88} it shows a
shift of the average value for the SiOSi angle from about 140$^{\circ}$
for the melt at 3600\,K to about 150$^{\circ}$ for the glass structure
(note the very good agreement with the value of 151$^{\circ}$, obtained
by an analysis of NMR data by Mauri {\it et al} \cite{mauri00}). In
contrast to that the BKS potential shows only a slight change of the
SiOSi angle from the melt to the glass structure \cite{vollmayr96}.

\subsection{Dynamic Features}
\begin{figure}
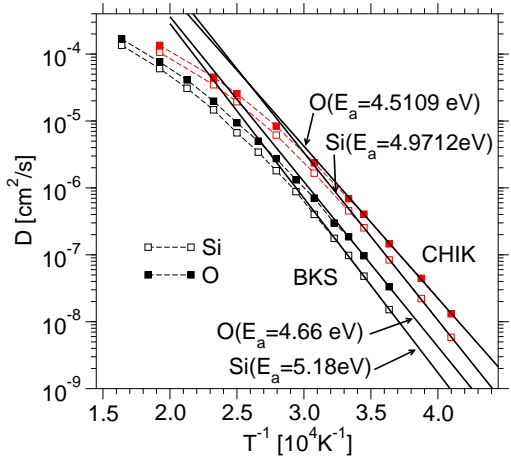

\onefigure[scale=0.33]{fig4.eps}
\caption{Arrhenius plot of the self-diffusion constants obtained by 
simulations with the BKS and the CHIK potentials. The bold 
solid lines are fits with Arrhenius laws (see text).}
\label{fig4}
\end{figure}
The self-diffusion constants $D_{\alpha}$ ($\alpha = {\rm Si,O}$)
were computed from the long-time limit of the mean squared
displacements $\langle r^2_{\alpha}(t)\rangle$ via the Einstein
relation $D_{\alpha}=\lim_{t\to\infty}\langle r^2_{\alpha}(t)\rangle/6t$
\cite{glassbook,allen}.  In Fig. \ref{fig4}, the temperature dependence of
the $D_{\alpha}$ is displayed in an Arrhenius plot.
The diffusion dynamics as predicted by the CHIK model appears to be faster
than that of the BKS model. At low temperatures ($T\simeq 2750$\,K),
the self-diffusion coefficients are about a factor of 5 higher than
those obtained from the BKS model.

In agreement with various experimental studies \cite{mazurin}, at low
temperatures the self-diffusion constants can be well-described by an
Arrhenius law, $D_{\alpha}=A_{\alpha} \exp[-E_{\rm a}^{\alpha}/(k_B T)]$.
The activation energies $E_{\rm a}$, that we find for the CHIK model,
are 4.51\,eV for oxygen and 4.97\,eV for silicon. These values are very
similar to those obtained for the BKS model \cite{horbach99} and they
are in good agreement with experimental results ($E_{\rm a}=4.7$\,eV
for oxygen \cite{mikkelsen84} and $E_{\rm a}=6.0$\,eV for silicon
\cite{brebec80}).

\begin{figure}
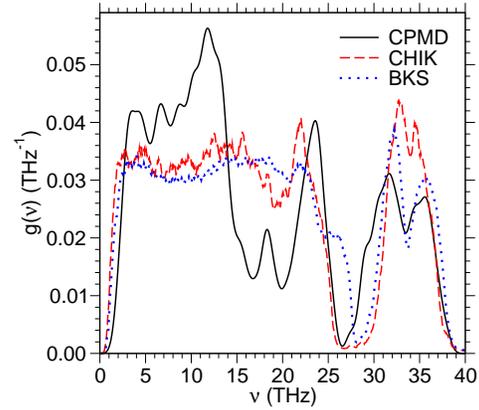

\onefigure[scale=0.30]{fig5.eps}
\caption{Density of states from the different MD simulations, as indicated.}
\label{fig5}
\end{figure}
To compute the vibrational density of states $g(\nu)$ (with $\nu$ the
frequency), 16 independent and fully equilibrated samples at 2440\,K
were quenched instantaneously to 300\,K, followed by an annealing
for 80\,ps. Then, $g(\nu)$ was determined by calculating the Fourier
transform of the velocity autocorrelation functions for Si and O (for details 
see Ref.~\cite{horbach99b}). In Fig.~\ref{fig5}, the so-obtained density of
states is compared to results from CPMD (from Ref.~\cite{Benoit_Kob02})
and MD simulations using the BKS potential (from Ref.~\cite{horbach99b}).

A prominent feature in $g(\nu)$ is the double-peak  in the
frequency band above about 28\,THz. The vibrational excitations
in this band correspond to stretching modes of the Si-O bond
\cite{glassbook}. Note that the BKS potential has been fitted to
reproduce the high frequency band of the vibrational spectrum.  Thus,
it is not surprising that it is in better agreement with CPMD than the
CHIK model, since we have not included any vibrational properties in
the fitting procedure of the CHIK potential. The CHIK model
seems to be better than the BKS model in the intermediate frequency
band 20\,THz$\le \nu \le$ 30\,THz where, in contrast to the BKS case,
a single peak is observed, albeit at a slightly lower frequency than
in CPMD result.  Below 20\,THz the BKS and CHIK results are
very similar and do not agree well with the CPMD result. In order to
significantly improve the description of the density of states it might
be necessary to account for polarization effects in the model potential,
as suggested by Wilson {\it et al.}~\cite{wilson93}.

\subsection{$\alpha-$quartz}
\begin{largetable}
\begin{tabular}{|c|c|c|c|c|c|} \hline
&  &  \multicolumn{2}{|c|}{BKS} &  \multicolumn{2}{|c|}{CHIK} \\  
\cline{3-4}\cline{4-6} \rb{Parameters} & \rb{Exp. \cite{will88}}& value &
error \% &  value &  error \%\\ \hline \hline
$V ({\rm \AA}^3$) & 112.933 & 115.2  & 2.0 & 121.6  & 7.7 \\ 
$a(=b) \, ({\rm\AA})$ & 4.91239 & 4.940 & 0.6 & 5.045 & 2.7 \\
$c ({\rm \AA}) $    & 5.40385 & 5.448 & 0.8 & 5.520 & 2.1 \\ 
$u$           & 0.4701  & 0.4648   & 1.1 & 0.4732   & 0.6 \\
$x$           & 0.4139  & 0.4268   & 3.1 & 0.4268   & 3.1 \\
$y$           & 0.2674  & 0.2715   & 1.5 & 0.2597   & 2.8 \\
$z$           & 0.2144  & 0.2085   & 2.7 & 0.2012   & 6.1 \\ \hline
\end{tabular}
\caption[]{Lattice parameters for $\alpha$-quartz for the BKS and the CHIK 
model, in comparison to experiment. \label{LEM_AQ_GR}}
\end{largetable}
\begin{largetable}
\begin{tabular}{|c|c|c|c|c|c|c|c|} \hline
Elastic & & \multicolumn{2}{|c|}{BKS} & \multicolumn{2}{|c|}{CHIK} \\ 
\cline{3-6} Constant (GPa) & \rb{Exp. \cite{mcskimin65,levien80}} & value & 
error \%  & value & error \% \\ \hline \hline                         
$C_{11}$ & 86.8  &  90.6 & 4.2  & 91.9  & 5.8   \\
$C_{33}$ & 105.8 & 107.0 & 1.1   & 91.6  & -13.4 \\
$C_{44}$ & 58.2  &  50.2 & -13.7 & 46.9  & -19.4 \\
$C_{66}$ & 39.9  &  41.2 & 3.2   & 42.6  & 6.7   \\
$C_{12}$ & 7.0   &   8.1 & 15.7  & 6.6   & -5.7  \\
$C_{13}$ & 19.1  &  15.2 & 20.4  & 18.5  & -3.1  \\
$C_{14}$ & -18.0 & -17.6 & 2.2   & -14.4 & 20.0  \\ \hline
\end{tabular}
\caption[]{Elastic constants for $\alpha-$quartz calculated at $0$ K for
different force fields, the experimental results are given for the room
temperature ($T=298$\,K). \label{Elastic_Constant}}
\end{largetable}
We have also carried out calculations on $\alpha-$quartz to check the
transferability of our potential.  In the experimental phase diagram
\cite{glassbook}, $\alpha-$quartz is the stable phase in the low
temperature domain (i.e.~up to 846\,K).  As already mentioned, some of
the crystalline properties of $\alpha-$quartz, i.e.~different lattice
parameters and elastic constants, have been used for the parametrization
of the BKS potential \cite{BKS}.  We have also calculated these quantities
using our potential (now using a cutoff at 10\,\AA~for the short-ranged
part of the potential). For this, the GULP code \cite{gulp} was employed
which yields the different lattice properties at 0\,K.  The results are
summarized in Tables~(\ref{LEM_AQ_GR}) and (\ref{Elastic_Constant}), in
comparison to those predicted by the BKS model and to experimental data
(the latter at 300\,K!).  The CHIK potential reproduces the experimental
data with comparable accuracy as the BKS potential, thus indicating that
it is also well suited for the study of crystalline SiO$_2$ phases.

\section{Conclusions}
This work was devoted to the development of a new effective potential
for silica. We have presented a fitting scheme for deriving
effective potentials from {\it ab initio} simulations (CPMD).
The new potential (named CHIK potential) is superior to the BKS model with
respect to various static and dynamic properties of amorphous silica.
The CHIK potential is also found to be transferable to other phases
such as $\alpha$-quartz, considering the quantitative agreement between
the experimental data for both cell parameters and elastic constants.
The fitting scheme proposed in this study is well-suited to
parametrize potentials for other (amorphous) systems, in particular
for mixtures of SiO$_2$ with other oxides such as alkali oxides or
Al$_2$O$_3$ where the local structure of the melt or the glass
can be very different from that of crystalline phases.

\acknowledgments
We are grateful to Kurt Binder for very valuable discussions.
We thank Schott-Glaswerke Mainz  for financial support through  
the Schott-Glaswerke-Fond and the
French Minister of National Education for his administrative founds
(co--tutelle). Computing time on the JUMP at the NIC J\"ulich, on the
CINES at Montpellier as well as the IDRIS at Orsay (France) are gratefully acknowledged.

\end{document}